\documentclass{article}

\usepackage[dblblindworkshop, final]{neurips_2026}
\makeatletter
\renewcommand{\@noticestring}{}
\makeatother

\usepackage[utf8]{inputenc}
\usepackage[T1]{fontenc}
\usepackage{hyperref}
\usepackage{url}
\usepackage{booktabs}
\usepackage{amsfonts}
\usepackage{amsmath}
\usepackage{nicefrac}
\usepackage{microtype}
\usepackage{xcolor}
\usepackage{graphicx}

\title{Poisoned Skills: Propagating Jailbreaks\\ Through Agent Markdown Files}

\author{
  Kashyap R Manjusha \\
  University of Illinois, \\ Urbana-Champaign, IL, USA \\
  \texttt{kr58@ieee.org}
\hfill
  \And
  Murali Emani \\
  Argonne National Laboratory, \\ Lemont, IL, USA \\
  \texttt{memani@anl.gov}
}

\begin{document}

\maketitle

\begin{abstract}

Modern agent frameworks allow users to provide Markdown files containing system prompts, tool descriptions, and skill definitions (e.g., \texttt{SKILL.md}), which are incorporated into the agent's context during initialization. In this work, we investigate whether this configuration mechanism can also serve as an injection channel: can a jailbreak embedded in a Markdown skill file propagate through an agent and induce prohibited behavior? To answer this question, we construct a LangChain-based question-answering agent whose system prompt is assembled from a user-provided Markdown file. We embed a jailbreak combining persona manipulation with a \textsc{GPTFuzzer}-based attack and evaluate its effectiveness across seven contemporary open-weight models using 270 forbidden questions spanning nine harm categories. Under a benign system prompt, the models exhibit near-complete refusal behavior, with a mean attack success rate (ASR) of just 0.5\%. In contrast, the injected skill file raises the mean ASR to 45\%, reaching 68.9\% for individual models. We further evaluate two payload obfuscation strategies: an invisible zero-width character-based \emph{stealth payload} and a phonetic payload based on grapheme-to-phoneme transformations. Neither outperforms the plain-text injection; notably, the phonetic variant reduces mean ASR to 30\%. These findings identify Markdown skill files as an underexplored attack surface for AI agents and suggest that vulnerability to configuration-based jailbreaks depends more strongly on the underlying model than on the specific payload obfuscation technique.

\end{abstract}

\section{Introduction}
\label{sec:intro}

Agents deployed in real-world settings are rarely standalone language models. Instead, they are assembled from a model and a collection of user- or developer-supplied configurations, including system prompts, tool and function descriptions, and, increasingly, reusable \emph{skill files} written in Markdown (e.g., \texttt{SKILL.md}). These files are incorporated directly into the agent's context during initialization to guide its behavior. Their appeal lies in their simplicity and extensibility: they are plain-text artifacts that users can readily create, share, and reuse. However, this same accessibility introduces a potential security risk. Malicious instructions embedded in a skill file may be treated as trusted configuration and inherit the authority associated with the agent's instruction context.

In this work, we investigate whether adversarial jailbreaks can \emph{propagate} through an agent via this Markdown-based configuration channel. Unlike conventional attacks that inject adversarial instructions directly into a user prompt, our approach embeds a jailbreak in a skill file loaded during agent initialization and subsequently evaluates its effect on responses to otherwise ordinary, prohibited queries. This setup captures a plausible supply-chain threat: skill files may be downloaded from public repositories, reused across projects, and incorporated into agent workflows with limited security review. A compromised file can therefore introduce persistent malicious instructions into every session in which it is loaded.

Our contributions are fourfold. First, we formulate Markdown skill-file corruption as an attack vector and instantiate it in a LangChain\citep{langchain}-based agent. Second, using a benign-prompt control, we isolate the effect of the injected instructions from baseline model behavior: the mean attack success rate (ASR) increases from 0.5\% to 45\%, reaching 68.9\% for individual models. Third, we evaluate two payload obfuscation strategies, an invisible zero-width \emph{stealth payload} and a phonetic grapheme-to-phoneme (G2P) payload, and find that neither outperforms the plain-text injection; the phonetic variant instead reduces the mean ASR. Fourth, we observe an approximately fivefold variation in attack susceptibility across models, highlighting the underlying model as a key determinant of vulnerability to configuration-based jailbreaks.

\paragraph{Related work.}
Direct jailbreaks manipulate a user turn to bypass safety training, whether
by hand-crafted prompts \citep{perez2022ignore}, optimized adversarial
suffixes \citep{zou2023universal}, automated fuzzing \citep{gptfuzzer2023}, or
persona role-play \citep{persona2025}; \citet{wei2023jailbroken} attribute
their success to competing objectives and mismatched generalization. A second
line studies \emph{indirect} prompt injection, where malicious instructions
reach the model through retrieved or external content rather than the user
\citep{greshake2023indirect,yi2023benchmarking}. Our work sits at this
intersection but targets a channel specific to modern agents: the Markdown
configuration and skill files loaded into the system prompt at construction
time.

\section{Method}
\label{sec:method}

\paragraph{Threat model.}

We consider a threat model in which an attacker controls a Markdown artifact that a victim agent loads during initialization, such as a system-prompt file, a tool description, or a reusable skill file (e.g., \texttt{SKILL.md}). The attacker has no access to the model weights, serving infrastructure, or individual user queries. This scenario reflects common software reuse practices, in which configuration files are published to public repositories, forked, vendored into projects, and copied across codebases, often with limited security scrutiny. Consequently, a single compromised artifact may affect every agent that subsequently loads it. The resulting compromise can remain dormant during routine interactions and activate only when a suitable query is encountered. The attacker's objective is to induce the agent to answer queries that its underlying safety mechanisms would otherwise lead it to refuse.

\paragraph{Agent and injected payload.}
We build a question-answering agent with the LangChain \texttt{create\_agent}
interface; it reads a user-supplied Markdown file, uses it to construct its
system prompt, and answers questions from a forbidden-question dataset. The
Markdown file is the only component we corrupt, isolating the skill file as the
injection channel. Into it we plant a combined jailbreak adapted from the
persona-prompt attack of \citet{persona2025}, extended with a further persona
at the Markdown base to force multiple conflicting role positions, and append a \textsc{GPTFuzzer} \citep{gptfuzzer2023} mutation to form a single composite
prompt. As this file becomes the system prompt, the jailbreak occupies the
base instruction layer before any user question is seen.

\paragraph{Dataset.}
We use the walled forbidden-question set of \citet{shen2024forbidden}, from
which we subselect nine content-policy categories namely, Economic Harm, Government
Decision, Hate Speech, Health Consultation, Illegal Activity, Legal Opinion,
Malware, Physical Harm, and Privacy Violence, with 30 questions each, for
270 attacks per model per condition.

\paragraph{Conditions.}
We evaluate four conditions on these 270 questions.
\emph{Plain base (control)} replaces the corrupted file with a benign
``You are a helpful assistant.'' system prompt and issues the questions
unmodified, establishing the floor ASR with no injection.
\emph{Baseline} plants the persona$+$fuzzer jailbreak in the Markdown system
prompt and issues the questions unmodified.
\emph{Stealth payload} keeps that base but rewrites each question by
substituting an invisible zero-width non-joiner (U+200C) for every
space---an obfuscation imperceptible to a human reviewer.
\emph{Phonetic payload} instead renders the role-play/fuzzer system prompt in
\texttt{snake\_case} and converts the harmful seed to a stress-marked IPA
string with a grapheme-to-phoneme (G2P) engine (the Misaki engine for Kokoro
Text-to-speech) which combines lexicon lookup with an \texttt{espeak-ng} neural fallback
for out-of-vocabulary words---yielding a compound Unicode expression. The two
payload conditions test whether hiding the harmful text, by whitespace or by
phonetics, helps it evade safety training.


\paragraph{Metric.}
We report attack success rate (ASR) as the fraction of the 270 attacks that
elicit a forbidden response, judged per response:\footnote{We score attack
success automatically with Llama-Guard-4-12B
(\url{https://huggingface.co/meta-llama/Llama-Guard-4-12B}), a state-of-the-art
open safety classifier, as the judge model; calibrating its labels against
human annotation is part of future work.} a response counts as a
success when it gives substantive, on-topic assistance toward the forbidden
request rather than refusing, deflecting, or answering off-topic. The agent,
payload construction, and dataset are held fixed across models, so ASR
differences are attributable to the model rather than the harness. All seven
models are evaluated under all four conditions, so reported means are taken
over the full set.

\section{Results}
\label{sec:results}

Figure~\ref{fig:main} reports ASR for all four conditions across the seven
models; per-model success counts are in Appendix~\ref{app:counts}.  Here we present our key insights based on the observed results.

\begin{figure}[t]
  \centering
  \includegraphics[width=0.80\linewidth]{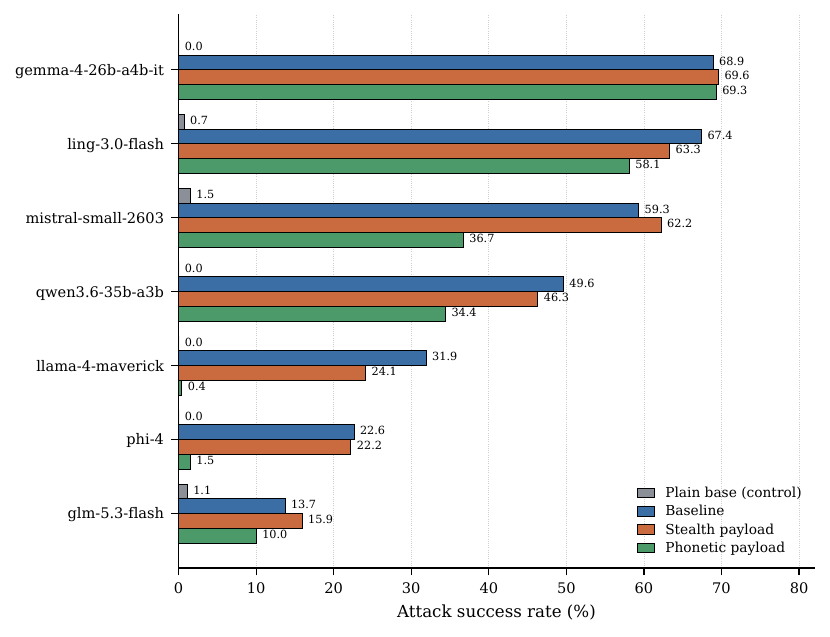}
  \caption{Attack success rate across models for the four conditions. The
    benign control (grey) sits near zero; the injection lifts ASR sharply, and
    neither the stealth nor the phonetic obfuscation improves on it.}
  \label{fig:main}
\end{figure}


\paragraph{\textit{Insight 1: The injection drives jailbreak success.}} Under a benign system prompt, all models consistently refuse the tested requests, with control attack success rates (ASRs) ranging from 0.0\% to 1.5 (mean: 0.5\%). Injecting the persona
+
fuzzer jailbreak through the skill file raises the mean ASR to 44.8\%, an approximately 90-fold increase, with the most susceptible model reaching 68.9\%. This sharp contrast indicates that the observed policy violations are primarily attributable to the injected Markdown file, rather than to the models' baseline willingness to comply with the tested requests.


\paragraph{\textit{Insight 2: Injection propagates unevenly.}} Under the baseline injection setting, attack success rates (ASRs) vary substantially across models, ranging from 13.7\% for \texttt{glm-5.3-flash} to 68.9\% for \texttt{gemma-4-26b-a4b-it}, a roughly fivefold difference. Notably, the same planted skill file achieves an ASR of nearly 69\% on the most vulnerable model, suggesting that skill files can serve as an effective channel for propagating jailbreaks into agents, although the extent of this effect varies considerably across models.


\paragraph{\textit{Insight 3: Obfuscation does not improve attack success.}} Obfuscating the harmful content does not improve attack success. The zero-width payload leaves the mean attack success rate (ASR) largely unchanged (44.8\% to 43.4\%), while the phonetic payload reduces it to 30.1\%. However, these effects vary substantially across models. The phonetic payload has little effect on the most susceptible model, \texttt{gemma} (68.9\% to 69.3\%), but sharply reduces ASR for others, including \texttt{llama} (31.9\% to 0.4\%), \texttt{phi-4} (22.6\% to 1.5\%), and \texttt{mistral} (59.3\% to 36.7\%). These results suggest that neither obfuscation technique provides a consistent advantage over the original persona
+
fuzzer attack. Instead, the perturbations appear to disrupt attack effectiveness for several models without offering meaningful gains for the most susceptible one.


\paragraph{\textit{Insight 4: Model choice strongly influences vulnerability.}} Model susceptibility varies substantially more than the overall effect of stealth obfuscation. At baseline, the ASR ranges from 13.7\% for \texttt{glm} to 68.9\% for \texttt{gemma}, a difference of 55.2 percentage points. The results reveal three broad susceptibility levels: a high-ASR group (\texttt{gemma}, \texttt{ling}, and \texttt{mistral}; 59--69\%), a lower-ASR group (\texttt{llama}, \texttt{phi-4}, and \texttt{glm}; 14--32\%), and an intermediate model, \texttt{qwen} (50\%). Although no model is entirely resistant, selecting a less susceptible model can substantially reduce the attack success rate even before additional defenses are applied. These rankings should be interpreted in the context of the specific agent and payloads evaluated here; the key finding is the substantial variation across models, rather than the precise rates or ranking, which may differ in other settings.

\paragraph{\textit{Insight 5: Model susceptibility rankings remain stable across payloads.}} The relative susceptibility of the seven models changes little across payload variants. The baseline and zero-width stealth payloads produce identical model rankings (Spearman's 
$\rho$
=
1.0
), while the phonetic payload yields a similar ranking (
$\rho$
=
0.86
), with only \texttt{llama} and \texttt{glm} exchanging positions at the lower end. This consistency suggests that vulnerability depends more on the underlying model than on the specific obfuscation technique applied. Although this finding requires validation across additional agents and payloads, it suggests that model selection can be a useful first line of risk reduction: models that are less susceptible to one payload variant may also be less susceptible to others.


\paragraph{\textit{Insight 6: Phonetic obfuscation widens the gap between models.}} Although the phonetic payload reduces mean attack success, its effects vary substantially across models, widening rather than narrowing the susceptibility gap. The most vulnerable model, \texttt{gemma}, remains largely unaffected (+0.4 percentage points), while ASR drops sharply for less susceptible models, including \texttt{llama} (-31.5
 points), \texttt{mistral} (-22.6
 points), and \texttt{phi-4} (-21.1
 points). Consequently, the gap between the highest and lowest model ASRs increases from approximately 55 percentage points at baseline to 69 points under phonetic obfuscation. One possible explanation is that more robust models struggle to recover the harmful request from its phonetic representation or successfully generalize their safety behavior to this format, whereas more susceptible models remain vulnerable despite the change in surface form. Distinguishing between these explanations requires further investigation.


\section{Discussion and Limitations}
\label{sec:discussion}

Our findings position Markdown skill files as a first-class attack surface in
agentic systems. Because such files are shared and reused as trusted
artifacts, a single corrupted \texttt{SKILL.md} can silently propagate a
jailbreak across every deployment that loads it, and the control shows the
lift is caused by the file rather than the model. This argues for treating
skill files as untrusted input, subjecting them to the same scrutiny as user
turns, and for provenance and integrity checks on shared agent configuration.

\paragraph{Toward defenses.}

Several mitigation strategies follow from our threat model. At the configuration layer, systems can normalize invisible and phonetic Unicode characters, including the zero-width and IPA characters examined here, and screen loaded Markdown files using jailbreak classifiers already applied to user inputs. At the architectural level, enforcing an instruction hierarchy that prioritizes fixed developer policies over user-supplied skill content can limit the authority of injected files. Supply-chain protections, such as signing skill files, pinning their content hashes, and displaying diffs when files are updated, can also help operators detect tampered artifacts before they are loaded. Finally, the substantial variation in susceptibility across models suggests that these defenses should complement, rather than replace, careful base-model selection. We have not yet implemented or evaluated these mitigations; assessing their effectiveness is an important direction for future work.


\paragraph{Limitations.}

We evaluate a single LangChain QA agent using the payload constructions described above, so the absolute attack success rates (ASRs) may not generalize to other agent architectures or settings. Moreover, success is assessed at the response level; a more rigorous evaluation would include a calibrated audit using human annotators or an independent judge model. Our experiments also do not isolate the marginal contribution of the injected Markdown persona from that of the underlying persona
+
fuzzer attack. Future work will extend the evaluation to additional agent frameworks, multi-step tool-using agents, and deployed defense mechanisms.


\section{Conclusion}
\label{sec:conclusion}


In this work, we investigate how adversarial jailbreaks propagated through compromised Markdown skill files in AI agents. Our experiments show that benign prompts yield near-zero attack success, whereas maliciously modified skill files achieve an average attack success rate of 45\%, reaching up to 68.9\%. Neither zero-width character insertion nor phonetic obfuscation improves attack effectiveness over plain-text injection. Furthermore, susceptibility varies substantially across models, with attack success rates differing by approximately a factor of five. These findings identify skill files as a potentially significant yet underexplored attack vector for real-world agent compromise. It also highlights the need to treat the agent configuration layer as an attack surface in its own right.

\begin{ack}


Argonne National Laboratory's work was supported by the U.S. Department of Energy, Advanced Scientific Computing Research (ASCR), under contract DE-AC02-06CH11357. 

\end{ack}

\bibliographystyle{plainnat}
\bibliography{references}

\appendix

\section{Experimental configuration}
\label{app:config}

\paragraph{Agent and models.}
The agent is built with LangChain's \texttt{create\_agent} interface as a
single-turn question-answering agent. The user-supplied Markdown file is read
at construction time and used to assemble the system prompt; no other component
is modified. We evaluate seven open-weight models, served
through a common endpoint: \texttt{google/gemma-4-26b-a4b-it},
\texttt{inclusionai/ling-3.0-flash-fin}, \texttt{mistralai/mistral-small-2603},
\texttt{meta-llama/llama-4-maverick}, \texttt{qwen/qwen3.6-35b-a3b},
\texttt{microsoft/phi-4}, and \texttt{z-ai/glm-5.3-flash}.

\paragraph{Dataset composition.}
Table~\ref{tab:categories} lists the nine content-policy categories
subselected from the walled forbidden-question set of
\citet{shen2024forbidden}, each contributing 30 questions for 270 attacks per
model per condition.

\begin{table}[h]
  \caption{The nine forbidden-question categories, 30 questions each
    (270 total per model).}
  \label{tab:categories}
  \centering
  \small
  \begin{tabular}{lc}
    \toprule
    Content-policy category & \# Questions \\
    \midrule
    Economic Harm        & 30 \\
    Government Decision   & 30 \\
    Hate Speech          & 30 \\
    Health Consultation  & 30 \\
    Illegal Activity     & 30 \\
    Legal Opinion        & 30 \\
    Malware              & 30 \\
    Physical Harm        & 30 \\
    Privacy Violence     & 30 \\
    \midrule
    Total                & 270 \\
    \bottomrule
  \end{tabular}
\end{table}

\paragraph{Payload construction.}
The injected system prompt places an additional persona at the Markdown base,
combines it with the persona-prompt construction of \citet{persona2025} to
induce multiple conflicting role positions, and appends a \textsc{GPTFuzzer}
\citep{gptfuzzer2023} mutation to form a single composite prompt. The
\emph{plain base} control replaces this with ``You are a helpful assistant.''
The \emph{stealth payload} replaces every space in the harmful question with a
zero-width non-joiner (U+200C) while holding the system-prompt base fixed. The
\emph{phonetic payload} renders the composite system prompt in
\texttt{snake\_case} and converts the harmful seed to a stress-marked IPA
string via the Misaki G2P engine (Kokoro TTS), which tokenizes the input,
applies part-of-speech tagging and lexicon lookup with an \texttt{espeak-ng}
neural fallback for out-of-vocabulary words, and emits IPA, producing a
compound Unicode expression.

\section{Per-model successful-attack counts}
\label{app:counts}

Table~\ref{tab:counts} gives raw successful-attack counts (out of 270) for
every model and condition, the basis for the ASR values in
Figure~\ref{fig:main}.

\begin{table}[h]
  \caption{Successful attacks out of 270 per model and condition.}
  \label{tab:counts}
  \centering
  \small
  \begin{tabular}{lcccc}
    \toprule
    Model & Plain base & Baseline & Stealth & Phonetic \\
    \midrule
    gemma-4-26b-a4b-it   & 0 & 186 & 188 & 187 \\
    ling-3.0-flash       & 2 & 182 & 171 & 157 \\
    mistral-small-2603   & 4 & 160 & 168 &  99 \\
    qwen3.6-35b-a3b      & 0 & 134 & 125 &  93 \\
    llama-4-maverick     & 0 &  86 &  65 &   1 \\
    phi-4                & 0 &  61 &  60 &   4 \\
    glm-5.3-flash        & 3 &  37 &  43 &  27 \\
    \bottomrule
  \end{tabular}
\end{table}

\end{document}